\documentclass{elsart}
\usepackage{graphicx,amssymb}

\begin{document}

\begin{frontmatter}

\title{Roundoff-induced attractors and reversibility in conservative two-dimensional maps}

\author{Guiomar Ruiz} 
\ead{guiomar@cbpf.br}  
 
\address{Centro Brasileiro de Pesquisas Fisicas\\
Rua Xavier Sigaud 150, 22290-180 Rio de Janeiro -- RJ, Brazil \\ 
and\\
Depto. Matem\'atica Aplicada y   Estad\'\i stica, EUIT
  Aeron\'auticos, UPM\\
Pza. Cardenal Cisneros n.4, E-28040 Madrid, Spain}
  
\author{Constantino Tsallis} 
\ead{tsallis@cbpf.br}  
\address{Centro Brasileiro de Pesquisas Fisicas\\
Rua Xavier Sigaud 150, 22290-180 Rio de Janeiro -- RJ, Brazil}

\date{\today}

\begin{abstract}

We numerically study two conservative two-dimensional maps, namely the baker map (whose Lyapunov exponent is known to be positive), and a typical one (exhibiting a vanishing Lyapunov exponent) chosen from the generalized shift family of maps introduced by C. Moore [Phys Rev 
Lett {\bf 64}, 2354 (1990)] in the context of undecidability. We calculated the time evolution of the entropy $S_q \equiv \frac{1-\sum_{i=1}^Wp_i^q}{q-1}$ ($S_1=S_{BG}\equiv -\sum_{i=1}^Wp_i \ln p_i$), and exhibited the dramatic effect introduced by numerical precision. 
Indeed, in spite of being area-preserving maps, they 
present, {\it well after} the initially concentrated ensemble has spread virtually all over the phase space, unexpected {\it pseudo-attractors} (fixed-point like for the baker map, and more complex structures for the Moore map). These pseudo-attractors, and the apparent time (partial) reversibility they provoke, gradually disappear for increasingly large precision. In the case of the Moore map, they are related to zero 
Lebesgue-measure effects associated with the frontiers existing in the definition of the map. 
In addition to the above, and consistently with the results by V. Latora and M. Baranger [Phys. Rev. Lett. {\bf 82}, 520 (1999)], we find that the rate of the far-from-equilibrium entropy production of baker map, numerically coincides with the standard Kolmogorov-Sinai entropy of this strongly chaotic system. 
\end{abstract}

\end{frontmatter}
\section{Introduction}
Dynamical systems which present exponential sensitivity to the initial conditions are called chaotic ({\it also called strongly chaotic}). Indeed, almost all orbits are unpredictable for any finite precision calculation, even if the evolution was purely deterministic. But dynamical systems at the edge of chaos are also somehow unpredictable, they typically exhibit a power-law sensitivity to the initial conditions, and are called {\it weakly chaotic}. For both cases, the sensitivity $\xi$ to the initial conditions is typically given \cite{tsallischaos,costa,robledo} by
\begin{equation}
\label{sensib}
\xi(t) \equiv \lim _{\Delta {\bf x}(0)\rightarrow 0} \frac{|\Delta {\bf x}(t)|}{|\Delta {\bf x}(0)|}= [1+(1-q)\lambda _q t]^{1/(1-q)}
\end{equation}
where $\Delta {\bf x}(t)$ is the time-dependent discrepancy between two initially close trajectories. The coefficient $\lambda _q$ is a generalized Lyapunov exponent, and $q$ is an index associated with the entropy \cite{Extensive} 
\begin{equation}
S_q \equiv \frac{1-\sum_{i=1}^Wp_i^q}{q-1} \,\,\,\, (q \in {\cal R}; \, S_1=S_{BG}\equiv -\sum_{i=1}^Wp_i \ln p_i) \,,
\label{qentropy}
\end{equation}
where $BG$ stands for {\it Boltzmann-Gibbs}. This entropy is at the basis of {\it nonextensive statistical mechanics}, which has received many applications for complex systems. Eq. (\ref{sensib}) recovers, in the $q=1$ limit, the usual exponential divergence $\xi=e^{\lambda_1 t}$ ($\lambda_1$ being, say for the simple one-dimensional case, the standard Lyapunov exponent). If $q=0$, we obtain the simple result $\xi \propto t$. In general, $q \ne 1$ yields a power-law dependence.

A meaningful statistical description is possible even when the maximal Lyapunov exponent vanishes. This fact has been illustrated in the Casati-Prosen triangle map \cite{casati}, a mixing and ergodic conservative system which presents the extreme case of linear instability \cite{casati}. This map satisfies that, in the infinitely fine graining limit (i.e., $W\to\infty$), the $q$-entropy increases linearly with time only for the value of the entropic index $q=0$ \cite{casatitsallis}; its slope is expected to coincide with the $q$-generalized Kolmogorov-Sinai entropy rate $\kappa_q$  \cite{tsallischaos}. Furthermore, this value is expected to coincide with the $q$-generalized Lyapunov coefficient; in other terms, a Pesin-like  equality \cite{pesin} is expected to hold for generic $q$ as well.

Conservative dynamical systems leading to entropic indices $q \ne 1$ are certainly interesting. The case that we have just mentioned, namely the Casati-Prosen map (for which $q=0$), is one such example. In the present paper we study various aspects along this direction, possibly with $q$ different from both unity and zero.

We are interested in studying the {\it entropy production}, i.e., the rate of increase of the $q$-entropy, of two-dimensional maps and, if possible, to connect it with the Kolmogorov-Sinai entropy rate \cite{kolmogorov} $\kappa_1$, which is a property defined for (strongly) chaotic dynamical systems. In this paper we present numerical results for the non-disipative baker map, and for one (from now on referred to as {\it Moore map}) of the shift-like dynamical systems proposed by Moore \cite{moore}. The latter is  expected to present a power-law sensitivity to the initial conditions, i.e., $q<1$ (possibly $q \ne 0$, in contrast with the Casati-Prosen map).  

To evaluate the $q$-entropy of the system, we first partition the phase space into $W>>1$ little equal cells, we then choose one of these cells, and put within $N>>1$ random initial conditions. As time $t$ evolves, the $N$ points spread over the phase space in such a way that, at each time $t$, we have a set of numbers $\{ N_i (t)\}$ ($\sum_{i=1}^{W}N_i(t)=N,\, \forall t$), so that $N_i(t)$ is the number of points inside the $i$-th cell. Then, for each value of $t$, we can consider a set of probabilities $\{p_i(t) \equiv N_i(t)/N\}$ to find $N_i$ points in the  $i$-th cell. For achieving a numerically meaningful definition of the probabilities $p_i$, the condition $N>>W$ has to be fulfilled; we typically consider $N=10 W$. At $t=0$, all probabilities but one are zero, hence $S_q(0)$, calculated through Eq. (\ref{qentropy}), vanishes $\forall q$. In other words, before the system starts to evolve, we know all the information regarding the occupancy of the phase space. As $t$ evolves, information is lost and $S_q(t)$ starts to increase. In all cases, $S_q$ will be bounded by its corresponding equiprobability value $ (S_q)_{max}=\frac{W^{1-q}-1}{1-q}$, whose $q=1$ limit case yields $\ln{W}$. 
The $q$-entropy production is then defined as
\begin{equation}
K_q\equiv \lim_{t\rightarrow \infty}  \lim_{W\rightarrow \infty}  \lim_{N\rightarrow \infty} \frac{S_q(t)}{t} \,.
\end{equation}
In practice, we take values of $(N,W,t)$ large enough so that our numerical result for $K_q$ becomes independent from them.

Summarizing, we numerically study direct snapshots of the occupancy of the space phase, and the time evolution of $S_q(t)$. From these, we shall present that

(i) $K_1=\kappa_1$ for the baker map;
 
(ii) numerical precision plays a relevant role in the phase space occupancy and time evolution of $S_q(t)$ in both baker and Moore maps;
 
(iii) the time evolution of $S_q$ partially (but not completely) reflects the time evolution of the phase space occupancy in both baker and Moore maps;
 
(iv) $q\ne 1 $ for the Moore map.

\section{The conservative baker map}
\begin{figure}[h]
\begin{center}
\hspace{0.33cm}\includegraphics[width=15.5cm,angle=0]{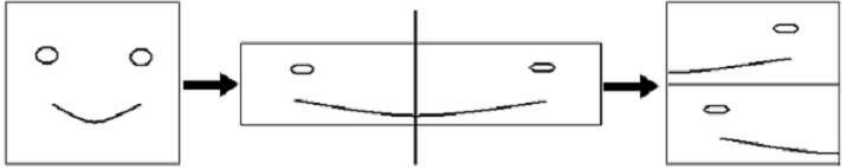}
\end{center}
\vspace{0.5cm}
\caption{\small The non-disipative baker map.}
\label{bakermap}
\end{figure}
The non-disipative baker map is an area-preserving 
chaotic map from the unit square into itself \cite{prigogine}. 
The procedure (see Fig. 1) is equivalent to the following transformation:
\begin{equation}
\left(x_{t+1},y_{t+1}\right) = 
\left\{
\begin{array}{ll}
\left(2x_t,y_t/2\right) & (0\le x_t <1/2) \\
\left(2x_t-1,(y_t+1)/2\right)  & (1/2 \le x_t \le 1),
\end{array}
\right.
\label{baker}
\end{equation}
where $x_{t}$ and $y_{t}$ are the coordinates of each point on step $t$, evolving to  $x_{t+1}$ and $y_{t+1}$ on step $t+1$.
Baker map is a mathematical model of a mixing system. It exhibits a strong sensitivity dependence on initial conditions, thanks to the stretching in the $x$-direction. Its two Lyapunov exponents ($\lambda_1=\ln{2}$, $\lambda_2=-\ln{2}$) are opposite as required by the preserving area condition. 

To study the q-entropy production we first asume that phase space is partitioned into a grid of $W=100 \times 100$ equal cells, and randomly choose $N=10^5$ random initial conditions inside one randomly chosen cell. As time evolves the points spread and, taking account the set of probabilities into each cell, we calculate the $q$-entropy as a function of time $t$ for different entropic indices $q$.    
In Fig. 2, we show the results we obtained working with double precision. Observe a first stage during which the $q$-entropies grow, and a second stage during which the $q$-entropies remain practically constant (at the value $S_q= \ln_q W \equiv \frac{W^{1-q}}{1-q}$, with $S_1=\ln W$) for a relatively long time. After that, during a third stage, the $q$-entropies decrease, eventually vanishing (the fact that the entropy vanishes means that all points are now within a single cell, which by no means necessarily coincides with the initially occupied cell). It looks like if they followed a decreasing law virtually identical to the temporal inversion of the first-stage increasing law: the map presents some sort of time reversibility. 
\begin{figure}[h]
\begin{center}\vspace{0.5cm}
\includegraphics[width=10.8cm,angle=0]{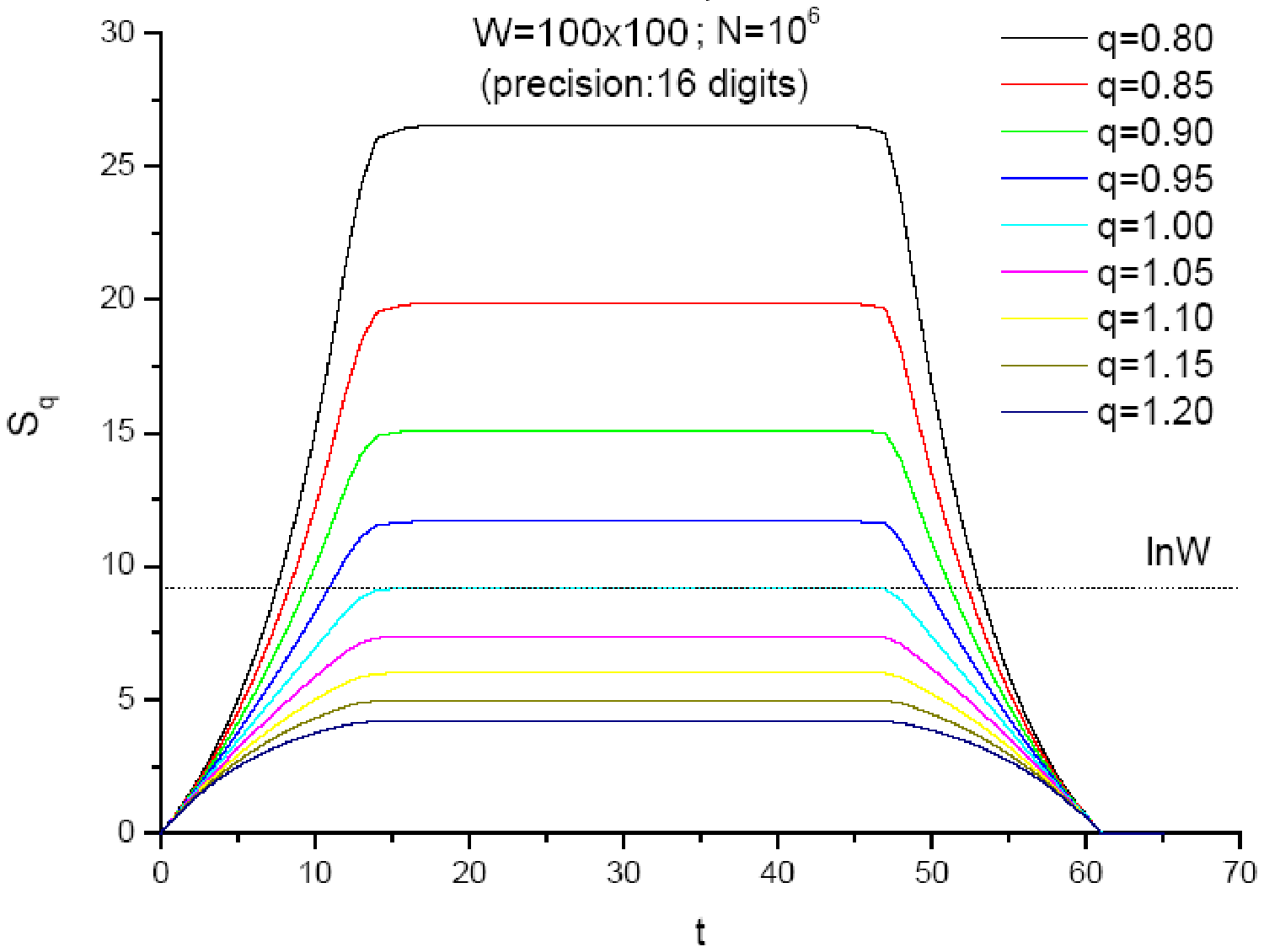}\\
\vspace{1.2cm}
\includegraphics[width=10.8cm,angle=0]{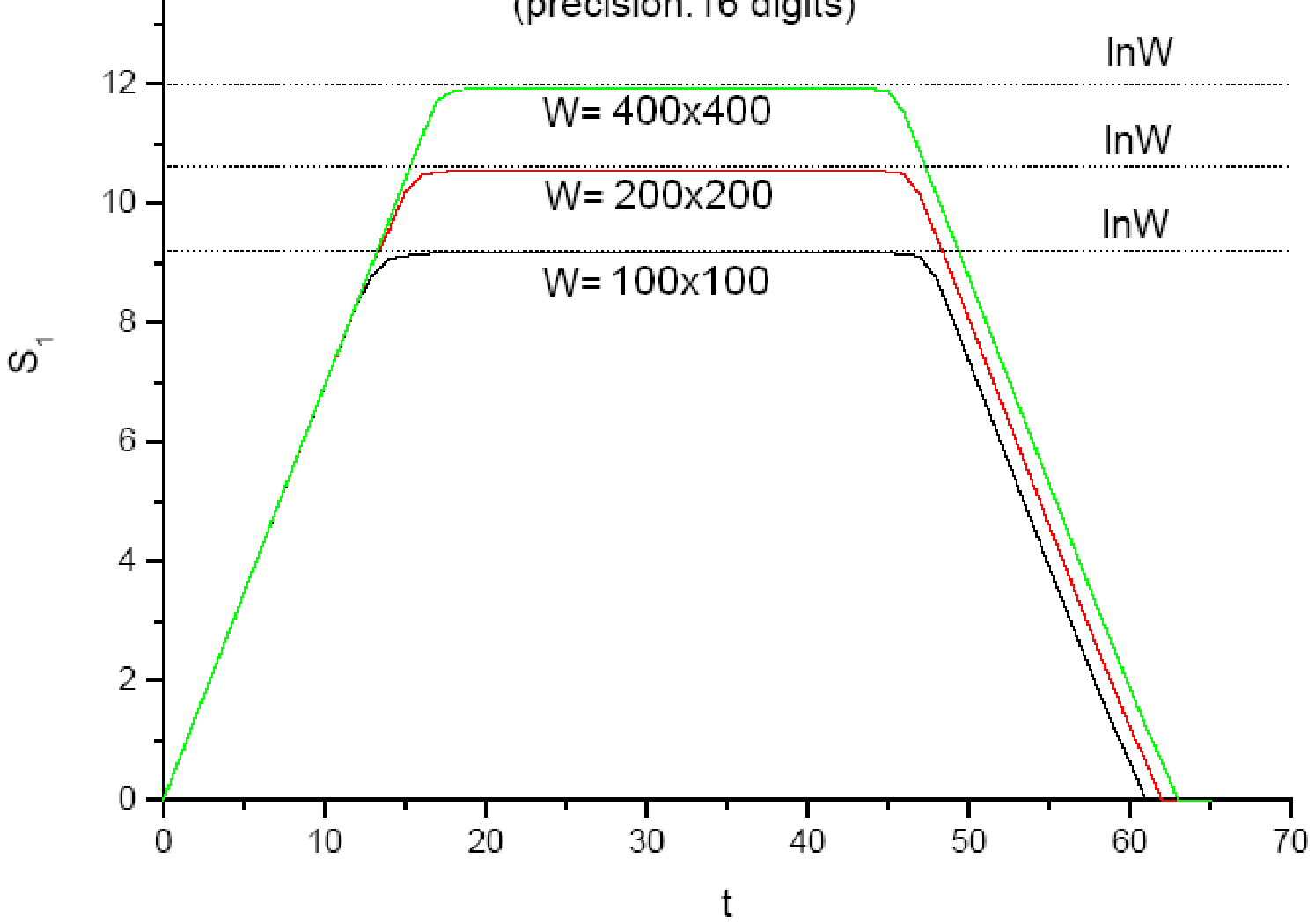}
\caption{\small
Time evolution of the $q$-entropy for the non-disipative baker map, using 16 digit calculations. The top figure shows $S_q(t)$ for entropic indices $q=0.80$, $0.85$, $0.90$, $0.95$, $1.00$, $1.05$, $1.10$, $1.15$, $1.20$ (from top to bottom) when $W=10^4$ and $N=10^6$. The bottom figure shows the results corresponding to $q=1$, for typical values of $W$ ($W=10^4$, $4\times 10^4$, $16\times 10^4$; $N=10W$). Notice that the bounding value for the $q=1$ entropy corresponds, in all cases, to equiprobability, i.e., $\ln W$. }
\label{bakerq}
\end{center}
\end{figure}
If we analyze the first stage, we observe that the value of $q=1$ corresponds to a {\it finite} $q$-entropy production per unit time. For  $q<1$ the curve is convex, while for $q>1$ it is concave. This is due to the fact that, in chaotic systems like baker map, i.e., with strong (exponential) sensitivity to the initial conditions, the usual Boltzmann-Gibbs entropy ($q=1$) is the appropriate one.
Fixing this value of the entropic index, we consider series of thinner partitions in phase space. The resulting $S_1$ entropy is shown in Fig. 2 for different values of $W$. 
Observe that, increasing $W$, the first stage remains for a longer time, which assures that the rate of $q$-entropy production for this far-from-equilibrium evolution depends neither on the number $N$ of points of the initial ensemble that we used, nor on $W$. 
If we perform a least-squares fit on these data, we obtain a linear regression index $R=0.99999998$. The slope of this linear rise is $0.6928\pm 0.0002$, which, with great precision, coincides with $\ln{2}$, the positive Lyapunov exponent. We conclude that the rate of $S_1$ production coincides, even for a finite size of coarse graining cells, with the standard Kolmogorov-Sinai entropy rate $\kappa_1$, which, according to Pesin theorem, equals the positive Lyapunov exponent. In other words, we verify
\begin{equation}
\kappa_1=K_1=\lambda_{1}\,.
\end{equation}
This result reinforces those  of Latora and Baranger \cite{latora,baranger}, in spite of the criticism of \cite{vulpiani}.
Indeed, Latora and Baranger showed, for several chaotic conservative dynamical systems, that the entropy evolution in far-from-equilibrium processes includes a stage during which the entropy linearly increases with time,  the slope being numerically equal to the standard Kolmogorov-Sinai entropy rate. All these results are consistent with the possible existence of a $q$-generalized theorem along the lines of the Pesin identity.

\begin{figure}[h]
\begin{center}\vspace{1.0cm}
\includegraphics[width=12cm,angle=0]{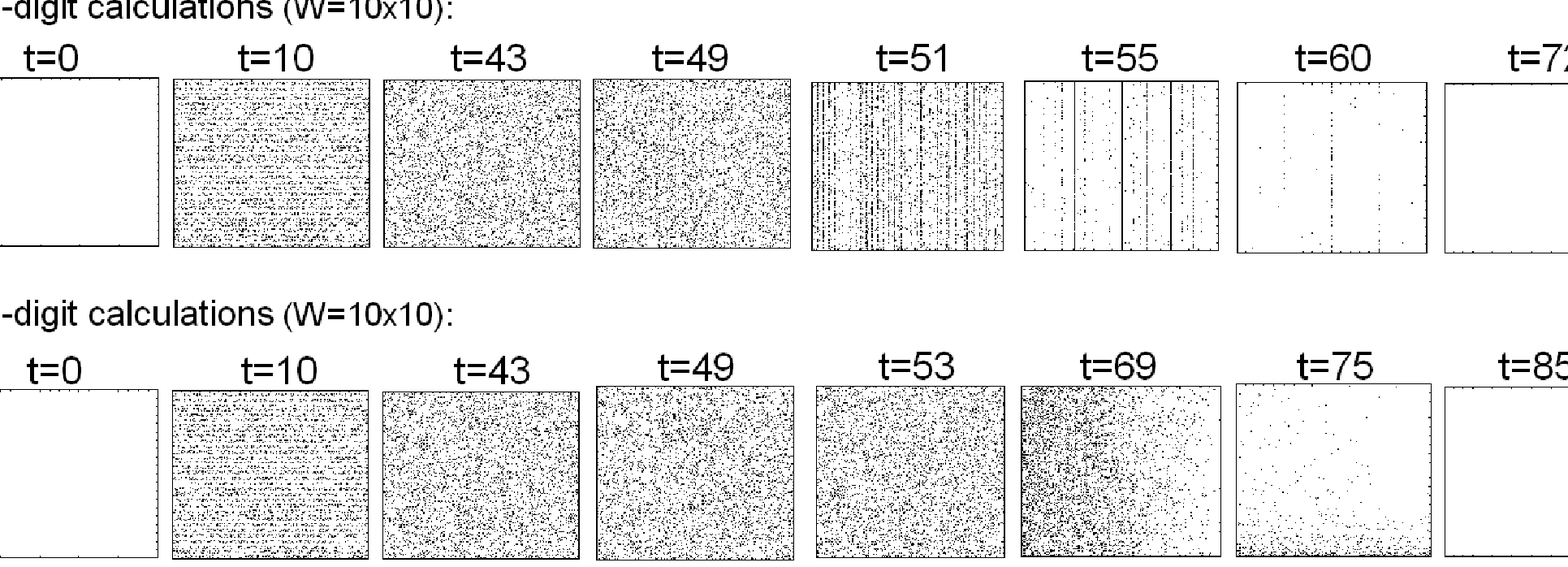}\\
\hspace{5.53cm}
\includegraphics[width=9cm,angle=0]{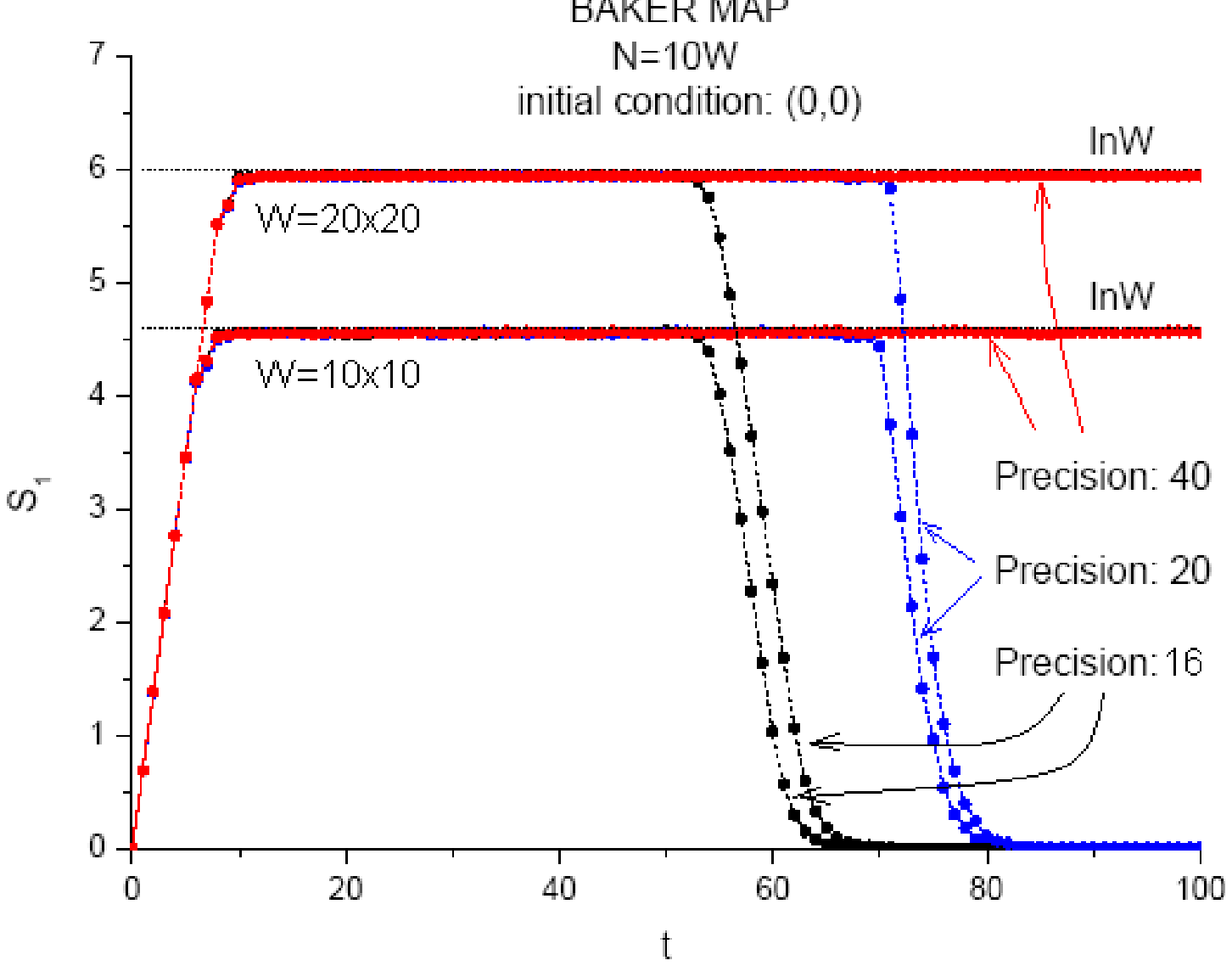}
\caption{\small  Numerical study of the baker map, with controlled fixed precision. The sequences of the top figure exhibit the evolution in phase space with a fixed precision. The corresponding curves for $S_1(t)$ are shown in the bottom figure. The evolution of $S_1(t)$ corresponding to a higher fixed precision experiment (40 digits )is shown as well; the time reversal of the entropy is not observed before $t=100$.}
\label{bakerq}
\end{center}
\end{figure}

The decreasing behavior of the $q$-entropy during the third stage needs further explanations. The decrease of entropy seemed to reflect a time reversal in the dynamic evolution. To elucidate this point, we performed iterations with controlled precision. The sequence of images in Fig. 3 shows that this reversal effect is gradually lost when precision is increased. Indeed, a roundoff-induced non-ergodic behavior emerges along the coordinate corresponding to the direction which exhibits strong sensitivity to the initial conditions. As a consequence, the $q$-entropy starts to decrease. This non-ergodic behavior eventually leads to a final collapse onto the fixed point $(0,0)$, which behaves as a pseudo-atractor. Consistently, the entropy turns back to be zero. Similarly, one must be aware that, on the plateau (i.e., the second stage), the entropy remains practically constant, in spite of the fact that it is possible to detect, in the evolution in phase space,  gradual changes of phase space occupancies (see the $W=100$ illustration in Fig. 3).
If we  increase the precision, we observe that the second stage is longer. In other words, the system presents an equal-probability occupation for a longer time. We conclude that the entropy decrease hopefully disappears at the infinite-precision limit. This observation has an important practical consequence, namely that special attention has to be systematically dedicated to possible precision effects. This is analog to the results of Longa et al. \cite{curado}, who revealed the importance of the precision in order to understand the phenomenon of coalescence of the trajectories in a chaotic system. 
\section{The (conservative) Moore map}
\begin{figure}[h]
\begin{center}
\hspace{1cm}\includegraphics[width=9.5cm,angle=0]{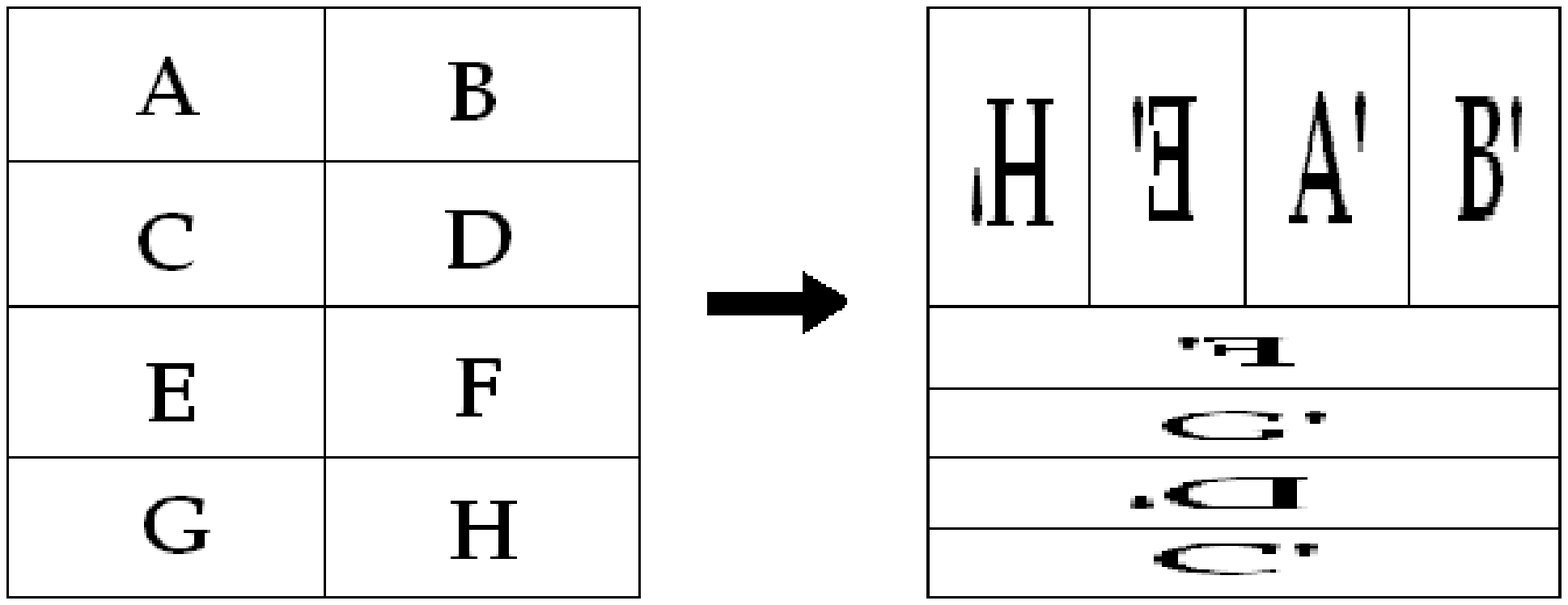}
\end{center}
\caption{\small The Moore map. Alphabetic symbols are written on the cells to show how local dynamics evolves}
\label{bakerq}
\end{figure}
\begin{figure}
\begin{center}
\vspace{5.5cm}\hspace{1.5cm}\includegraphics[width=9cm,angle=0]{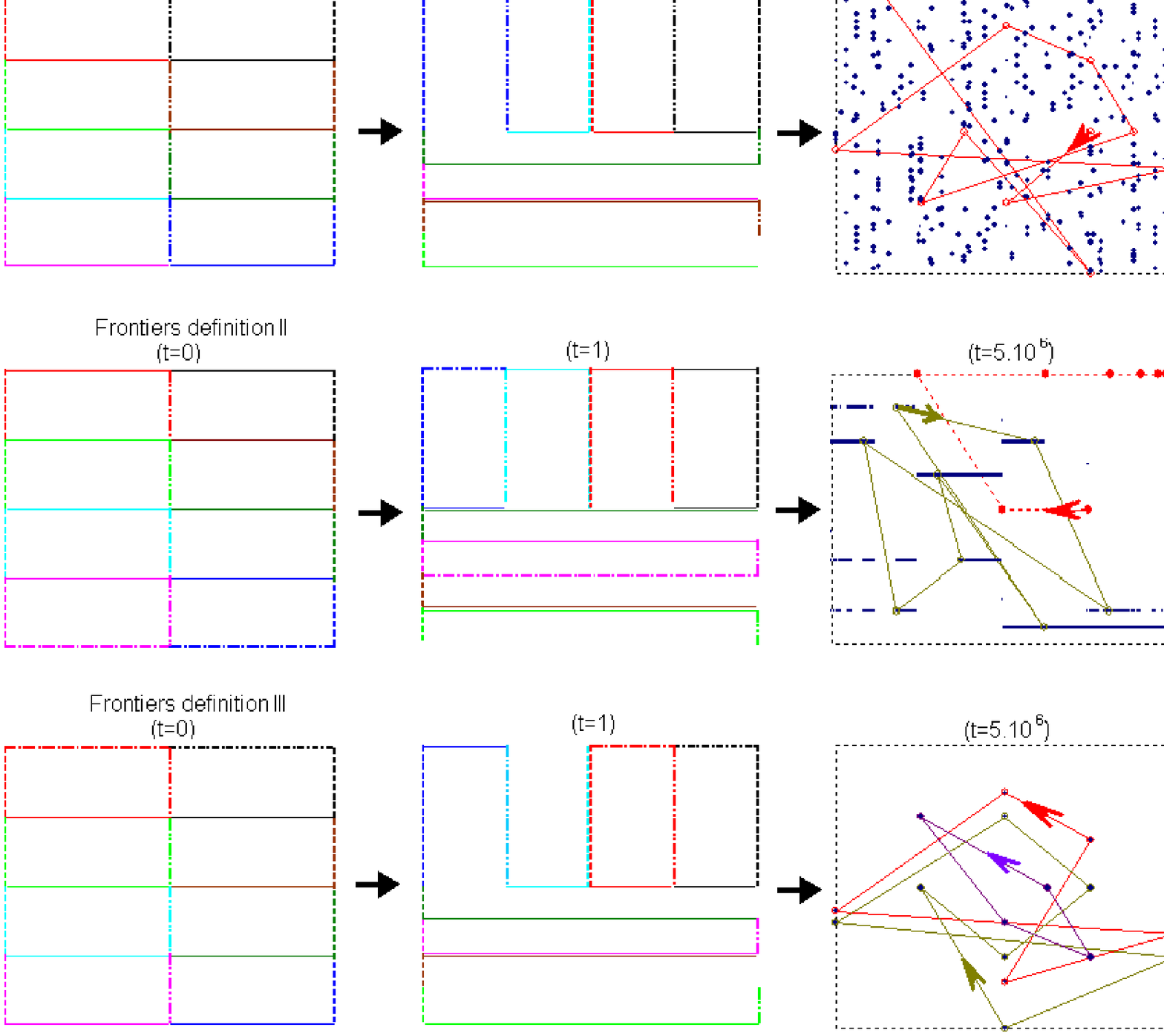}\\
\vspace{1cm}\hspace{0.6cm}
\includegraphics[width=6.5cm,angle=0]{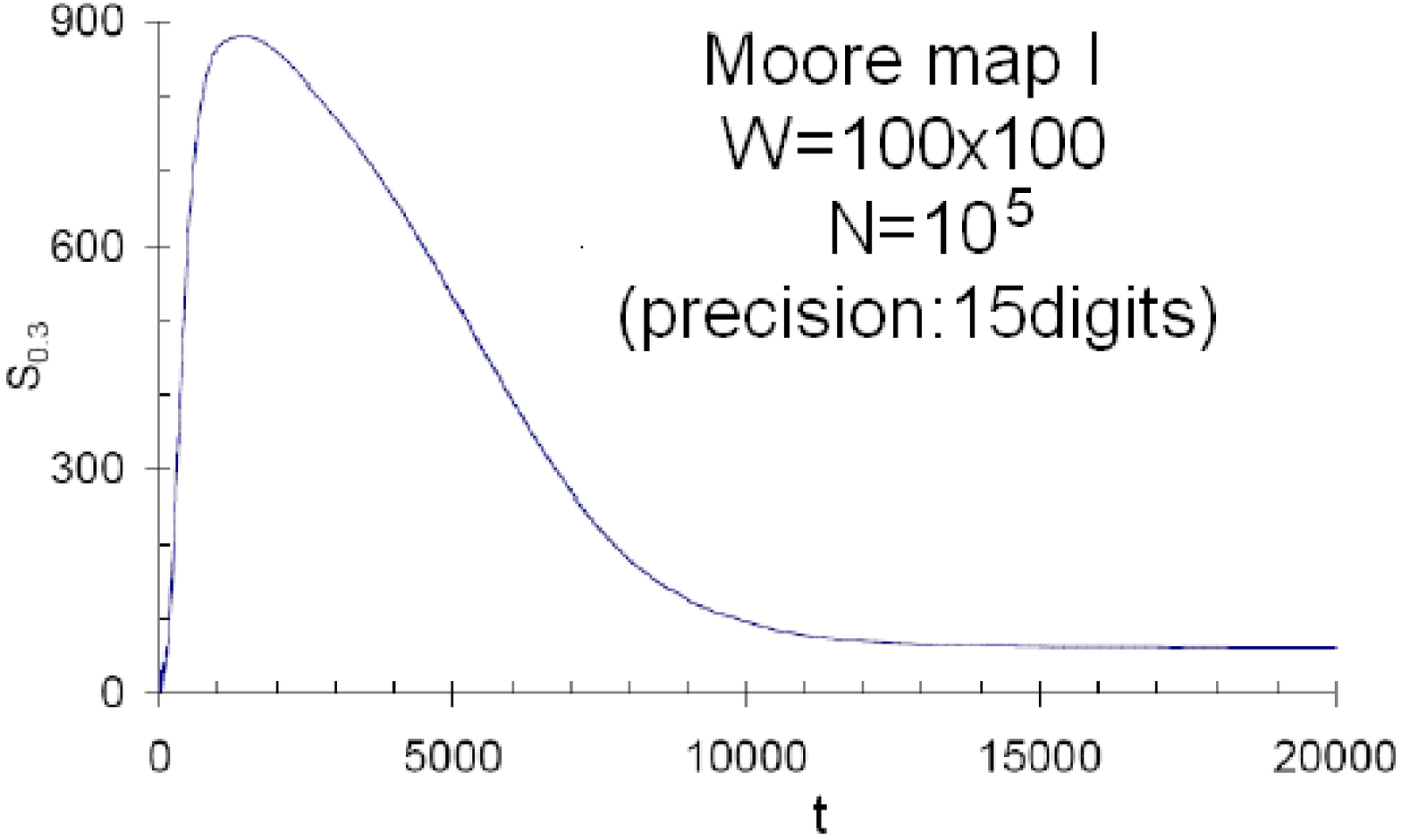}
\includegraphics[width=6.5cm,angle=0]{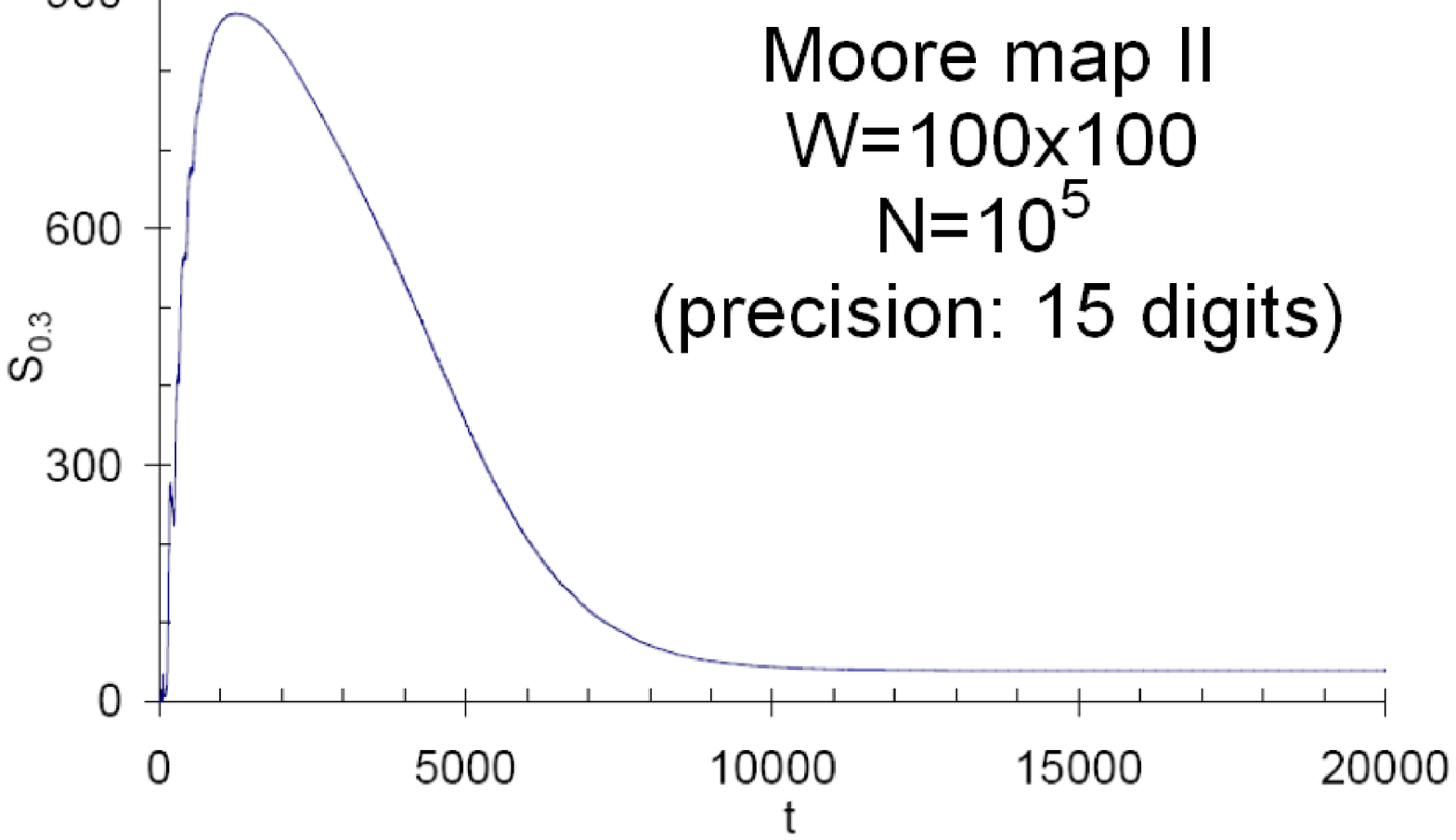}
\caption{\small Time evolution of three Moore maps (denoted by I, II and III) which differ just in the definition of the mapping of the frontiers.
{\it Top figures:} Snapshots of the evolution in phase space ($t=0$, $t=1$ and $t=5 \times 10^6$), when starting with points exclusively {\it on} the frontiers. In the $t=5 \times 10^6$ squares, we have also indicated typical trajectories. {\it Bottom figures:} Time evolution of $S_{0.3}$ for  maps I and II, starting with a set of initial conditions within a small cell  ($W=10^4$; $N=10^5$).}
\label{bakerq}
\end{center}
\end{figure}

The generalized shift family of maps proposed by Moore \cite{moore} are a class of dynamical systems that posses some sort of undecidability, as compared with other low--dimensional chaotic systems \cite{moore,moore2}. 
We are interested in studing a paradigmatic one within the Moore family. It is equivalent to the piecewise linear map shown in Fig. 4. When this map is recurrently applied, the area in phase space is conserved, while the corresponding shape keeps changing in time, becoming increasingly complicated. 
This map appears to be ergodic, possibly exhibits a Lyapunov exponent $\lambda_1=0$, and, presumably, the divergence of close initial conditions follows a power-law behavior \cite{moore}. When we consider a partition of $W$ equal cells and select $N$ random initial conditions inside one random cell, the points spread much slower than they do on  the baker map. More precisely, they spread, through a slow relaxation process, all over the phase space, each orbit appearing to gradually fill up the entire square.
\begin{figure}
\begin{center}\vspace{5.5cm}\hspace{1cm}
\includegraphics[width=10.5cm,angle=0]{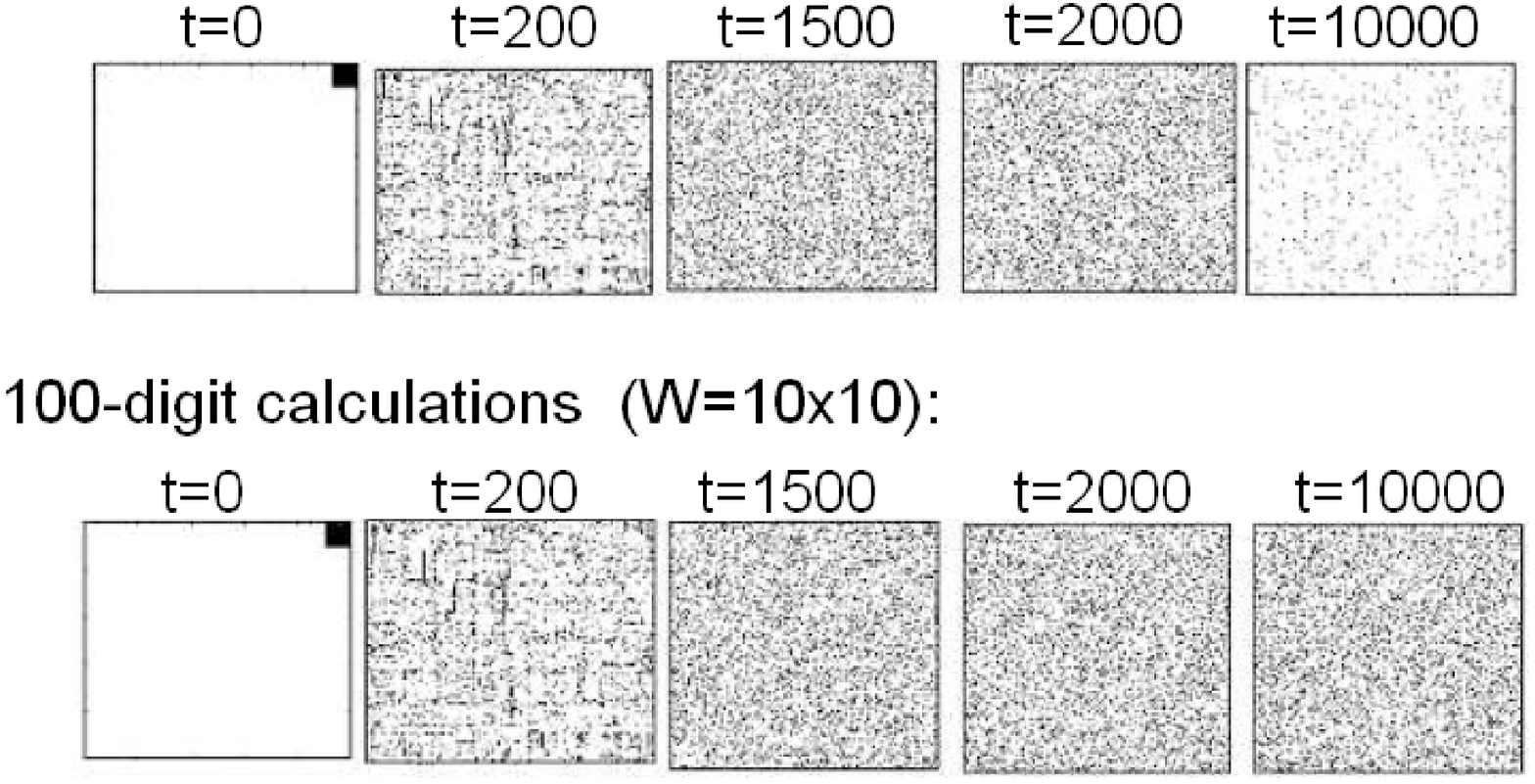} \\ \vspace{1cm}
\includegraphics[width=15cm,angle=0]{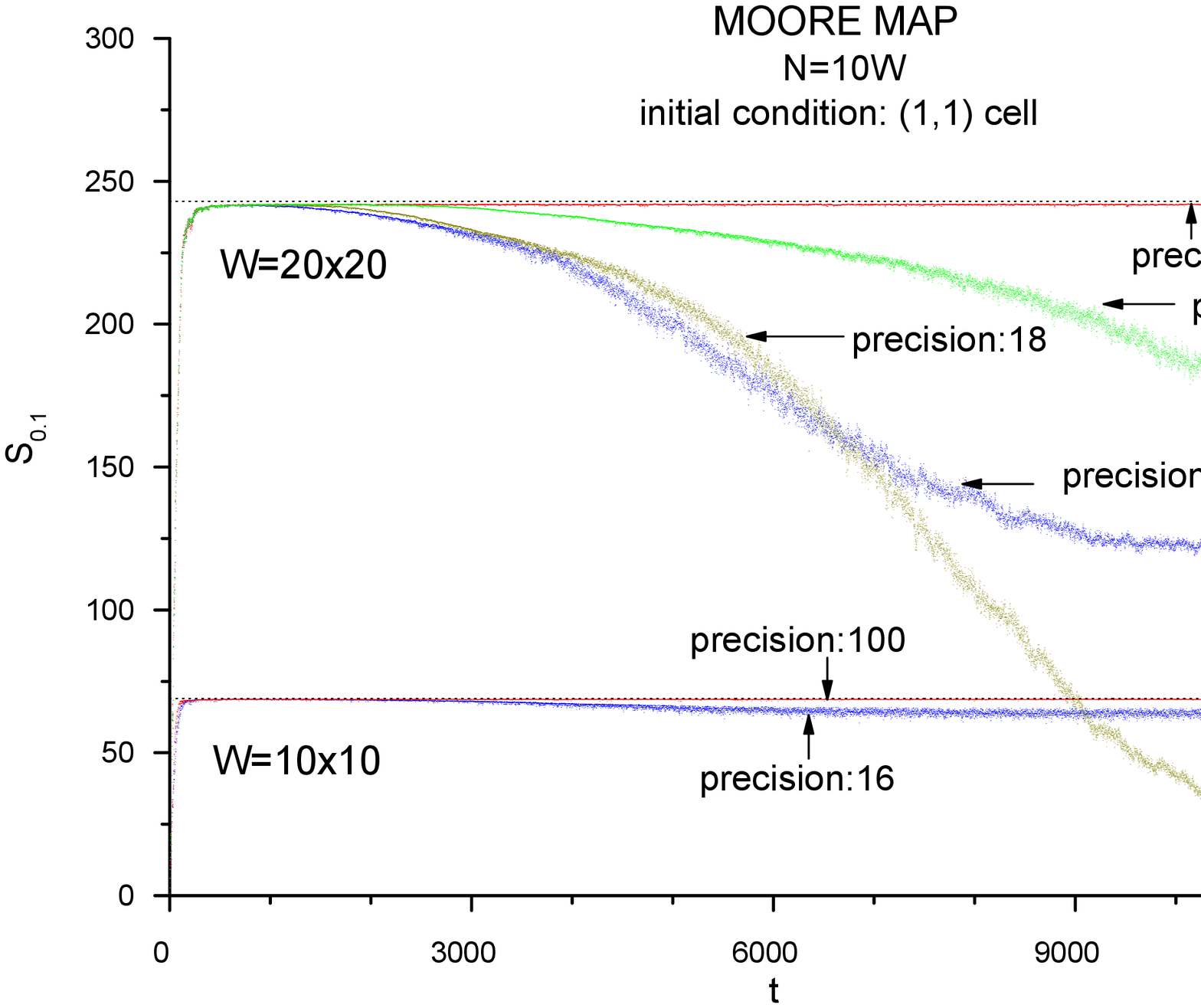}\vspace{0.3cm}
\caption{\small Numerical study of the Moore map I. 
{\it Top figures:} Evolution of occupancy in phase space. {\it Bottom figure:} Evolution of $S_{0.1}$.}
\vspace{0.53cm}
\label{bakerq}
\end{center}
\end{figure}

When we study the dynamic evolution corresponding to this map we observe that, in spite of being an area-preserving two-dimensional map, i. e. a {\it conservative} map, it presents a sort of unexpected {\it pseudo-attractors}. The three temporal sequences in Fig. 5, correspond to three maps which only differ in the definition of the mapping of the frontiers existing in the piecewise linear Moore map. Different  pseudo-attractors emerge for different mappings of the frontiers. The bottom figures reflect how the pseudo-attractors affect the long-time evolution of the respective $q$-entropies. If we define a map such that the area is preserved but the frontiers are dissipated, no pseudo-attractor emerges. This fact shows that the pseudo-attractors are due to zero Lebesgue-measure effects related to the frontiers. Consequently, in high-precision numerical studies, the roundoff-induced long-time non-ergodic behavior tends to disappear. 

On calculations, with say sixteen digits, we observe that $S_q$ initially grows. Then, its increasing rate slows down so as to tend towards the constant value that characterizes equiprobability. When precision is gradually lost along time, and non-ergodicity behavior emerges, $S_q$ starts decreasing. The influences of the precision and of the grid size are illustrated in Fig. 6.

\begin{figure}
\begin{center}
\vspace{1.5cm}
\includegraphics[width=9.5cm,angle=0]{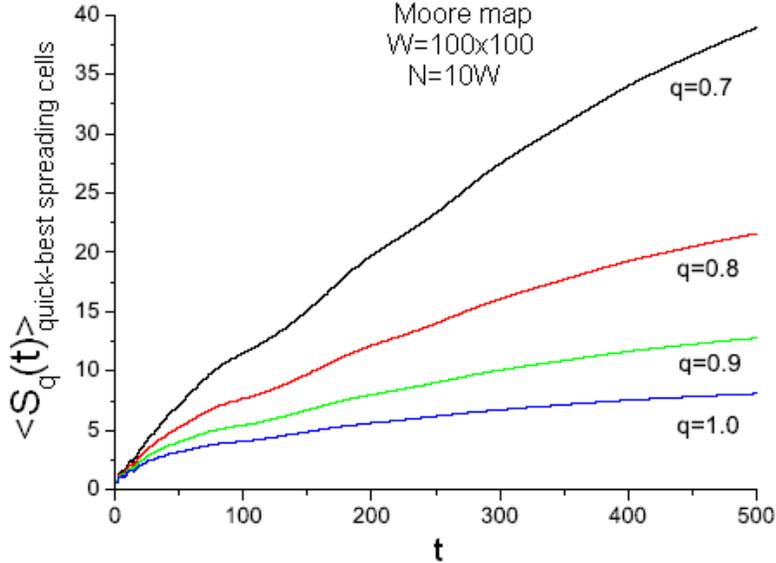}\vspace{0.5cm}
\end{center}
\caption{\small Time dependence of $S_q$ averaged over the $10 \%$ quick-best spreading cells, on the far-from-equilibrium regime, for typical values of $q$.}
\label{mooresq}\vspace{1.5cm}
\end{figure}
The study of the $q$-entropy growth on the far-from equilibrium regime reveals important fluctuations. Fluctuations are usual at chaos thresholds and they requiere efficient and careful averaging over the initial conditions \cite{latoraCT}. We checked that roundoff effects are not important in this regime and that they cancel over random averages. The optimal $q$-entropy average process consists in averaging over the quick-best spreading cells, which should be taken as independent initial conditions \cite{latoraCT}. The $q$-entropy averages over the $10 \%$ quick-best spreading cells presents, for all values of $q$, intrinsic oscillations (see Fig. 7) depending neither on averages nor on precision. We verify that $q=1$ does {\it not} produce, even discounting the intrinsic oscillations, a linear increase of $S_q(t)$. In other words, the usual Boltzmann-Gibbs entropy is not the appropriate one for simply describing this system. We consistently expect $q<1$. However, its precise value elluded us because the oscillations make the numerical analysis quite complex.  The determination of $q$ remains as a future goal.
\section{Main results}
Let us summarize our main results: 

(i) The coincidence between the Kolmogorov-Sinai entropy rate $\kappa_1$ and the entropy production $K_1$ is numerically shown on a far-from-equilibrium conservative chaotic system (baker map). 

(ii) Roundoff-induced pseudo-reversibility and pseudo-attractors are found on two dimensional chaotic conservative maps. 

(iii) A statistical entropy may only partially reflect the changes of distributions in the phase space for any finite coarse graining and any finite precision.

(iv) In order to have a {\it finite} entropy production for $S_q$ we need a value of $q$ which is definitively smaller than unity, i.e., the Boltzmann-Gibbs entropy does not appear as the most adequate tool. Deeper studies are needed in order to establish whether another value of $q$ can solve this problem.

As we see, the entropic issues related with time evolution for deterministic nonlinear dynamical systems that are conservative are not very different from those corresponding to the dissipative ones.
\section*{Acknowledgements}
We warmly thank interesting remarks by Alfredo M. O. Almeida, Andrea Rapisarda, Antonio Politi, Cristopher Moore, Evaldo M.F. Curado and  Raul O. Vallejos. One of us (G. R.) acknowledges financial support from  the Universidad Polit\'{e}cnica de Madrid. Partial financial support from Pronex/MCT, CNPq and Faperj (Brazilian Agencies) is acknowledged as well. 

\end{document}